\documentclass[prl,twocolumn,nofootinbib]{revtex4}
\usepackage[utf8]{inputenc}
\usepackage{graphicx}
\usepackage{amsmath,amssymb}
\usepackage{amsfonts}
\usepackage{physics}
\usepackage{xspace}
\usepackage{bm}
\usepackage{mathrsfs}
\usepackage{nicefrac}
\usepackage[dvipsnames]{xcolor}
\usepackage[unicode]{hyperref}
\hypersetup{colorlinks=true, citecolor=MidnightBlue, linkcolor=CornflowerBlue, urlcolor=CornflowerBlue, linktocpage=true}
\usepackage[all]{hypcap}
\usepackage{multirow}
\usepackage[format=plain,justification=RaggedRight,labelsep=endash,font=small,labelfont=sc]{caption}
\usepackage[export]{adjustbox}

\definecolor {darkgreen}{rgb}{0.2,0.7,0.2}
\newcommand{\pdagger}{{\phantom{\dagger}}}
 \newcommand{\psbar}{\bar{\psi}}

\begin{document}
\title{Pseudo-gauge dependence of quantum fluctuations of energy in a hot relativistic gas of fermions}

\author{Arpan Das$^{1}$}
\author{Wojciech Florkowski$^{2}$}
\author{Radoslaw Ryblewski${^1}$}
\author{Rajeev Singh${^1}$}

\affiliation{$^{1}$Institute  of  Nuclear  Physics  Polish  Academy  of  Sciences,  PL-31-342  Krak\'ow,  Poland}
\affiliation{$^{2}$Institute of Theoretical Physics, Jagiellonian University, PL-30-348 Krak\'ow, Poland}

\begin{abstract}
Explicit expressions for quantum fluctuations of energy in subsystems of a hot relativistic gas of spin-$\nicefrac{1}{2}$ particles are derived. The results depend on the form of the energy-momentum tensor used in the calculations, which is a feature described as pseudo-gauge dependence. However, for sufficiently large subsystems the results obtained in different pseudo-gauges converge and agree  with the canonical-ensemble formula known from statistical physics. As different forms of the energy-momentum tensor of a gas are a priori equivalent, our finding suggests that the concept of quantum fluctuations of energy in very small thermodynamic systems is pseudo-gauge dependent. On the practical side, the results of our calculations determine a scale of coarse graining for which the choice of the pseudo-gauge becomes irrelevant.
\end{abstract}

\date{\today \hspace{0.2truecm}}

\maketitle
\flushbottom

\par\textit{Introduction.---}
Quantum fluctuations arising from the quantum uncertainty relation and statistical fluctuations inherent to any many-body system~\cite{Huang:1987asp} play a very important role in physics as they encode the information about possible phase transitions~\cite{Smoluchowski,PhysRevLett.85.2076}, dissipative phenomena~\cite{Kubo1} or even the formation of structures in the Early Universe~\cite{Lifshitz:1963ps,PhysRevLett.49.1110}. 

Very recently,  we have analyzed the quantum fluctuations of energy in subsystems of a relativistic gas of bosons and demonstrated that they diverge for the subsystem size approaching zero, however, they agree with thermodynamic fluctuations in canonical ensemble if subsystems become sufficiently large~\cite{Das:2020ddr}. In this work, we investigate a complementary but at the same time significantly different aspect of quantum fluctuations of energy, namely, we study their dependence on the so-called pseudo-gauge transformation. The latter refers to our freedom in choosing a specific form of the energy-momentum tensor to describe the system's behavior. To be more explicit, for any original energy-momentum tensor $\hat{T}^{\mu\nu}$ satisfying the continuity equation $\partial_\mu \hat{T}^{\mu\nu}=0$ we can construct a different one by adding the divergence of an antisymmetric object, namely~\cite{Chen:2018cts,HEHL197655,Speranza:2020ilk}
\begin{equation}
\hat{T}^{\prime \,\mu\nu} = \hat{T}^{\mu\nu} + \partial_\lambda \hat{A}^{\nu\mu \lambda}
\label{eq:PG}
\end{equation}
with $\hat{A}^{\nu\mu \lambda} = - \hat{A}^{\nu\lambda \mu}$. By construction, the new tensor is also conserved, i.e.,  $\partial_\mu \hat{T}^{\prime \, \mu\nu}~=~0$.

To study the effects of different pseudo-gauges we consider a hot relativistic gas of spin-$\nicefrac{1}{2}$ particles. Besides the most popular canonical  version of the energy-momentum tensor that is obtained from the Noether theorem, we take into consideration the Belinfante-Rosenfeld version (BR)~\cite{BELINFANTE1939887,BELINFANTE1940449,Rosenfeld1940}, the de Groot-van Leeuwen-van  Weert  version (GLW)~\cite{DeGroot:1980dk}, and the Hilgevoord-Wouthuysen formulation (HW)~\cite{HILGEVOORD19631,HILGEVOORD19651002}. All these forms have been recently broadly discussed in the context of spin-polarized media~\cite{Florkowski:2018fap,Speranza:2020ilk}, as the pseudo-gauge transformation can be introduced also for the spin tensor $\hat{S}^{\lambda, \mu\nu}$ that is a part of the total angular momentum tensor $\hat{J}^{\lambda, \mu\nu} = \hat{L}^{\lambda, \mu\nu}+\hat{S}^{\lambda, \mu\nu}$~\cite{HEHL197655,Speranza:2020ilk,Leader:2013jra,Gallegos:2021bzp}.

Our main object of interest is the energy density operator defined as the time-time (``$tt$") component of the energy-momentum tensor, $\hat{T}^{00}$. Our results show that, although $\hat{T}^{00}$ does depend on the pseudo-gauge choice, the thermal average of such operator is independent of the pseudo-gauge. However, for small subsystems the fluctuation of $\hat{T}^{00}$ depends on the pseudo-gauge choice and only for sufficiently large systems such fluctuations become pseudo-gauge independent.

Our analysis sheds new light on the concept of energy density used in the classical description of fluids, in particular, in the context of the relativistic hydrodynamics of hot matter produced in relativistic heavy-ion collisions~\cite{Jaiswal:2016hex,Florkowski:2017olj,Romatschke:2017ejr}. As different forms of the energy-momentum tensor are fundamentally equivalent for the description of a gas, our finding suggests that the concept of quantum fluctuations of energy in very small thermodynamic systems has no absolute physical meaning and should be always considered in a specific pseudo-gauge chosen for the system's analysis. On the other hand, from the practical point of view, our calculations determine a typical scale of coarse graining for which the choice of the pseudo-gauge becomes irrelevant. Such a scale depends in the considered case on the temperature and mass of the particles that make up the gas. 

 \textit{Basic concepts and definitions.---} Following our previous study~\cite{Das:2020ddr}, in the present work we consider a subsystem $S_a$ of the thermodynamic system $S_V$ composed of spin-$\nicefrac{1}{2}$ particles with mass $m$ described by the canonical ensemble characterized by the temperature $T$ (or its inverse, $\beta = 1/T$). The characteristic volume of the subsystem $S_a$ is always smaller than the volume $V$ of the system $S_V$, and $V$ is sufficiently large to allow for performing integrals over particle momenta (otherwise one commonly introduces sums imposed by the box periodic conditions).\footnote{We use the West--Coast metric $g_{\mu\nu} = \hbox{diag}(+1,-1,-1,-1)$. Three-vectors are shown in bold font and a dot is used to denote the scalar product of both four- and three-vectors, i.e., $a^\mu b_\mu = a \cdot b = a^0 b^0 - \boldsymbol{a} \cdot  \boldsymbol{b}$.  } 

With these assumptions in mind, we describe our system by a spin-$\nicefrac{1}{2}$ field in thermal equilibrium. The field operator has the standard form~\cite{Tinti:2020gyh}
\begin{align}
\psi(t,\boldsymbol{x})=&\sum_r\int\frac{d^3k}{(2\pi)^3\sqrt{2\omega_{\boldsymbol{ k}}}}\Big(U_r^{\pdagger}(\boldsymbol{k})a_r^{\pdagger}(\boldsymbol{k})e^{-i k \cdot x} \nonumber\\
&~~~~~~~~~~~~~~~~~~~~~~~+V_r^{\pdagger}(\boldsymbol{k})b_r^{\dagger}(\boldsymbol{k})e^{i k \cdot x} \Big),
\label{equ1ver1}
\end{align}
where $a_r^{\pdagger}(\boldsymbol{k})$ and $b_r^{\dagger}(\boldsymbol{k})$ are annihilation and creation operators for particles and antiparticles, respectively. They satisfy the canonical anti-commutation relations, $\{a_r^{\pdagger}(\boldsymbol{k}),a_s^{\dagger}(\boldsymbol{k}^{\prime})\} =(2\pi)^3\delta_{rs} \delta^{(3)}(\boldsymbol{k}-\boldsymbol{k}^{\prime})$ and
$ \{b_r^{\pdagger}(\boldsymbol{k}),b_s^{\dagger}(\boldsymbol{k}^{\prime})\} =(2\pi)^3\delta_{rs} \delta^{(3)}(\boldsymbol{k}-\boldsymbol{k}^{\prime})$, while
all the other operators anticommute with each other. The index $r$ represents the polarization degree of freedom. The Dirac spinors $U_r^{\pdagger}(\boldsymbol{k})$ and $V_r^{\pdagger}(\boldsymbol{k})$ have normalization ${\bar U}_r^{\pdagger}(\boldsymbol{k}) U_s^{\pdagger}(\boldsymbol{k}) = 2 m \delta_{rs}$ and ${\bar V}_r^{\pdagger}(\boldsymbol{k}) V_s^{\pdagger}(\boldsymbol{k}) = -2 m \delta_{rs}$, and the quantity $\omega_{\boldsymbol{k}}=\sqrt{\boldsymbol{k}^2+m^2}$ is the energy of a particle. 

To perform thermal averaging, it is sufficient to know the expectation values of the products of two and four creation and/or annihilation operators (for both particles and antiparticles)  ~\cite{CohenTannoudji:422962,Itzykson:1980rh,Evans:1996bha}
\begin{align}
& \langle a_r^{\dagger}({\boldsymbol{k}})a_s^{\pdagger}({\boldsymbol{k}}^{\prime})\rangle=(2\pi)^3\delta_{rs}\delta^{(3)}({\boldsymbol{k}}-{\boldsymbol{k}}^{\prime})f(\omega_{\boldsymbol{k}}),\label{equ2ver1}\\
& \langle a^{\dagger}_r(\boldsymbol{k})a^{\dagger}_s(\boldsymbol{k}^{\prime})a_{r^{\prime}}^{\pdagger}(\boldsymbol{p})a_{s^{\prime}}^{\pdagger}(\boldsymbol{p}^{\prime})\rangle\nonumber\\
& =(2\pi)^6 \Big(\delta_{rs^{\prime}}\delta_{r^{\prime}s}\delta^{(3)}(\boldsymbol{k}-\boldsymbol{p}^{\prime})~\delta^{(3)}(\boldsymbol{k}^{\prime}-\boldsymbol{p})\nonumber\\
&-\delta_{rr^{\prime}}\delta_{ss^{\prime}}\delta^{(3)}({\boldsymbol{k}}-\boldsymbol{p})~\delta^{(3)}({\boldsymbol{k}}^{\prime}-\boldsymbol{p}^{\prime})\Big)f(\omega_{{\boldsymbol{k}}})f(\omega_{{\boldsymbol{k}}^{\prime}}).\label{equ4ver1}
\end{align}
Here $f(\omega_{{\boldsymbol{k}}})$ is the Fermi--Dirac distribution function for particles. For antiparticles, the Fermi--Dirac distribution function differs by the sign of the chemical potential $\mu$. However, in the absence of any conserved charge, as here, we have $\mu=0$. Any other combination of two and four creation and/or annihilation operators can be obtained from Eqs.~\eqref{equ2ver1} and \eqref{equ4ver1}
through the anticommutation relations for $a_r^{\pdagger}({\boldsymbol{k}})$, $a_r^{\dagger}({\boldsymbol{k}})$, 
$b_r^{\pdagger}({\boldsymbol{k}})$, and $b_r^{\dagger}({\boldsymbol{k}})$.

Following~\cite{Chen:2018cts}, we define an operator $\hat{T}^{00}_a$ that represents the energy density of a subsystem $S_a$ placed at the origin of the coordinate system~\cite{Chen:2018cts}
\begin{align}
\hat{T}^{00}_a = \frac{1}{(a\sqrt{\pi})^3}\int d^3\boldsymbol{x}~\hat{T}^{00}(x)~\exp\left(-\frac{{\boldsymbol{x}}^2}{a^2}\right).
\label{equ6ver1}
\end{align}
In Eq.~\eqref{equ6ver1} a smooth Gaussian profile has been used to define the subsystem $S_a$. Instead of a cube, the smooth profile with a length scale $a$ has been used to remove possible sharp-boundary effects. The thermal expectation value of the normal ordered operator $:\hat{T}^{00}_a:$ is denoted as $\langle :\hat{T}^{00}_a :\rangle$. To remove an infinite vacuum part coming from zero-point energy contributions, we apply standard normal ordering procedure to $\hat{T}^{00}_a$. To determine the fluctuation of the energy density of the subsystem $S_a$, we consider the variance
\begin{equation}
 \sigma^2(a,m,T) = \langle :\hat{T}^{00}_a: :\hat{T}^{00}_a: \rangle - \langle :\hat{T}^{00}_a :\rangle^2\, 
 \label{sigma2}
\end{equation}
and the normalized standard deviation 
\begin{equation} 
\sigma_n(a,m,T)= \frac{(\langle:\hat{T}^{00}_a::\hat{T}^{00}_a:\rangle- \langle :\hat{T}^{00}_a :\rangle^2)^{1/2}}{\langle :\hat{T}^{00}_a :\rangle}.
\end{equation}
According to the Noether theorem, for each continuous symmetry of the action, there exists a corresponding conserved current. In particular, the translational symmetry and Lorentz invariance lead to the conserved canonical energy-momentum and total angular momentum tensors, respectively.
However, different pairs of the energy-momentum and total angular momentum tensors can be obtained by either changing the Lagrangian density or using the pseudo-gauge transformation as described above by Eq.~(\ref{eq:PG})~\cite{HEHL197655,Speranza:2020ilk,Florkowski:2018fap}. 

Among such alternative forms, certain choices seem to be more physically appealing. For example, since the canonical energy-momentum tensor is not symmetric, one often uses a pseudo-gauge to replace it with a symmetric one. The most popular method for such a symmetrization is due to Belinfante and Rosenfeld~\cite{BELINFANTE1939887,BELINFANTE1940449,Rosenfeld1940}. Very often the argument for having a symmetric $\hat{T}^{\mu\nu}$ is given, which comes from general theory of relativity (GR). In this case the energy-momentum tensor is assumed to be symmetric because it is obtained as the variation of the action with respect to the space-time metric. This argument has been recently questioned, since $\hat{T}^{\mu\nu}$ appearing in GR is purely classical, while the asymmetric parts of $\hat{T}^{\mu\nu}$ are of quantum origin~\cite{Florkowski:2018fap}.

In this work different forms of the energy-momentum tensors which appear in the current literature are studied. On the general grounds we do not find any reasons to argue that one of them is more fundamental or better for the description of a relativistic gas. We start with the canonical form and then subsequently turn to BR, GLW, and HW versions.

\medskip
 \textit{Canonical framework.---}
The canonical energy-momentum tensor of a spin-$\nicefrac{1}{2}$ field is defined as
 \begin{align}
     \hat{T}^{\mu\nu}_{\text{Can}}=\frac{i}{2}\bar\psi\gamma^{\mu}\overleftrightarrow{\partial}^{\nu}\psi-g^{\mu\nu}\mathcal{L}_{\text{D}}=\frac{i}{2}\bar\psi\gamma^{\mu}\overleftrightarrow{\partial}^{\nu}\psi.
     \label{equ9ver1}
 \end{align}
Here $\mathcal{L}_{\text{D}}$ denotes the Lagrangian density of a spin-$\nicefrac{1}{2}$ field, which can be expressed as
\begin{align}
    \mathcal{L}_{\text{D}}=\frac{i}{2}\bar\psi\gamma^{\mu}\overleftrightarrow{\partial}_{\!\!\mu}\psi-m\bar{\psi}\psi,
\end{align}
where $\overleftrightarrow{\partial}^{\mu}\equiv\overrightarrow{\partial}^{\mu}-\overleftarrow{\partial}^{\mu}$. To achieve the second equality in Eq.~\eqref{equ9ver1} one uses the equation of motion for $\psi$ and $\bar{\psi}$, i.e., the Dirac equation~\cite{Tinti:2020gyh}. Using the ``$tt$" component of the canonical energy-momentum tensor as defined in Eq.~\eqref{equ9ver1}, the thermal expectation value of the operator $\hat{T}^{00}_{\text{Can},a}$ for the subsystem $S_a$, can be easily shown to be
\begin{align}
\langle:\hat{T}^{00}_{\text{Can},a}:\rangle
& = 4\int\frac{d^3k}{(2\pi)^3}~\omega_{\boldsymbol{k}}~f(\omega_{\boldsymbol{k}}) \equiv\varepsilon_{\text{Can}}(T). 
\label{equ11ver1}
\end{align}
In Eq.~\eqref{equ11ver1} the factor of $4$ includes the spin degeneracy factor ($g_s=(2s+1)$) and the particle-antiparticle doubling. We note that Eq.~\eqref{equ11ver1} agrees with the elementary kinetic theory considerations~\cite{Huang:1987asp}. 
We also note that the canonical energy density $\varepsilon_{\text{Can}}(T)$, as defined by Eq.~\eqref{equ11ver1}, is not only independent of time but also independent  of the system size $a$, reflecting the spatial  uniformity of the system. 

Using the thermal expectation values of the products of two and four creation and/or annihilation operators as given by Eqs.~\eqref{equ2ver1}--\eqref{equ4ver1}, we find the energy density fluctuation for the canonical energy-momentum tensor 
\begin{align}
&\sigma^2_{\text{Can}}(a,m,T) =  2\int dK ~dK^{\prime} f(\omega_{{\boldsymbol{k}}})(1-f(\omega_{{\boldsymbol{k}}^{\prime}}))\nonumber\\
&\times \bigg[(\omega_{{\boldsymbol{k}}}+\omega_{{\boldsymbol{k}}^{\prime}})^2(\omega_{{\boldsymbol{k}}}\omega_{{\boldsymbol{k}}^{\prime}}+{\boldsymbol{k}}\cdot{\boldsymbol{k}}^{\prime}+m^2)e^{-\frac{a^2}{2}({\boldsymbol{k}}-{\boldsymbol{k}}^{\prime})^2}\nonumber\\
&-(\omega_{{\boldsymbol{k}}}-\omega_{{\boldsymbol{k}}^{\prime}})^2(\omega_{{\boldsymbol{k}}}\omega_{{\boldsymbol{k}}^{\prime}}+{\boldsymbol{k}}\cdot{\boldsymbol{k}}^{\prime}-m^2)e^{-\frac{a^2}{2}({\boldsymbol{k}}+{\boldsymbol{k}}^{\prime})^2}\bigg],
\label{equ12ver1}
\end{align}
where $dK \equiv d^3{{k}}/((2\pi)^{3} 2 \omega_{{\boldsymbol{k}}})$. Although normal ordering removes unwanted vacuum contribution to the energy-momentum tensor, it does not remove all the vacuum divergences in all composite operators. Therefore, following the arguments presented in Ref.~\cite{Das:2020ddr} we discard a divergent temperature-independent vacuum term that originally appears in Eq.~\eqref{equ12ver1}. 

\smallskip
 \textit{Belinfante-Rosenfeld framework.---}
Starting with the canonical energy-momentum tensor given by Eq.~\eqref{equ9ver1} and the canonical spin tensor, and performing the Belinfante-Rosenfeld symmetrization, one obtains~\cite{Tinti:2020gyh}
\begin{align}
    \hat{T}^{\mu\nu}_\text{BR}=\frac{i}{2}\bar{\psi}\gamma^{\mu}\overleftrightarrow{\partial}^\nu\psi-\frac{i}{16}\partial_{\lambda}\Big(\bar{\psi}\Big\{\gamma^{\lambda},\Big[\gamma^{\mu},\gamma^{\nu}\Big]\Big\}\psi\Big).
\label{equ13ver1}
\end{align}
Although  the Belinfante-Rosenfeld energy-momentum tensor as given in Eq.~\eqref{equ13ver1} is not manifestly symmetric under $\mu\leftrightarrow \nu$ exchange, using the Euler-Lagrange equations of motion for the fields or using the angular momentum conservation for the canonical tensors, the symmetry of $\hat{T}^{\mu\nu}_\text{BR}$ under $\mu\leftrightarrow \nu$ exchange can be proved~\cite{Tinti:2020gyh}. From Eqs.~\eqref{equ9ver1} and \eqref{equ13ver1}  it is clear that ``$tt$" components agree, i.e., $\hat{T}^{00}_\text{BR} = \hat{T}^{00}_\text{Can}$. Therefore, thermal average of the normal ordered $\hat{T}^{00}_\text{BR}$ and its fluctuation will be the same as $\varepsilon_{\text{Can}}(T)$, see Eq.~\eqref{equ11ver1}, and $\sigma^2_{\text{Can}}(a,m,T)$, see Eq.~\eqref{equ12ver1}, respectively. 

\smallskip
\textit{de Groot-van~Leeuwen-van~Weert framework.---}
The de~Groot-van~Leeuwen-van~Weert symmetric energy-momentum tensor for a massive spin-$\nicefrac{1}{2}$ field has the following form~\cite{DeGroot:1980dk}
\begin{align}
    \hat{T}^{\mu\nu}_{\text{GLW}}&= -\frac{1}{4m}\bar{\psi}\overleftrightarrow{\partial}^{\mu}\overleftrightarrow{\partial}^{\nu}\psi-g^{\mu\nu}\mathcal{L}_{\text{D}}\nonumber\\
    & =\frac{1}{4m}\Big[-\bar{\psi}(\partial^{\mu}\partial^{\nu}\psi)+(\partial^{\mu}\bar{\psi})(\partial^{\nu}\psi) +(\partial^{\nu}\bar{\psi})(\partial^{\mu}\psi)\nonumber\\
    &~~~~~~~~~~~-(\partial^{\mu}\partial^{\nu}\bar{\psi})\psi\Big].
    \label{equ14ver1}
\end{align}
To obtain the second line in Eq.~\eqref{equ14ver1} we have used the Dirac equations for $\psi$ and $\bar{\psi}$. The GLW symmetric energy-momentum tensor given by Eq.~\eqref{equ14ver1} is manifestly symmetric under $\mu\leftrightarrow\nu$ exchange. Following the same procedure as we have used for the canonical energy-momentum tensor and using the thermal average of the two and four operators, as given by Eqs.~\eqref{equ2ver1}--\eqref{equ4ver1}, we obtain
    \begin{align}
\langle:\hat{T}^{00}_{\text{GLW},a}:\rangle
& = 4\int\frac{d^3k}{(2\pi)^3}~\omega_{\boldsymbol{k}}f(\omega_{\boldsymbol{k}}) \equiv\varepsilon_{\text{GLW}}(T)
\label{equ15ver1}
\end{align}
and  
\begin{align}
&\sigma^2_{\text{GLW}}(a,m,T) =  \frac{1}{2m^2}\int dK ~dK^{\prime} f(\omega_{{\boldsymbol{k}}})(1-f(\omega_{{\boldsymbol{k}}^{\prime}}))\nonumber\\
&\times \bigg[(\omega_{{\boldsymbol{k}}}+\omega_{{\boldsymbol{k}}^{\prime}})^4\left(\omega_{{\boldsymbol{k}}}\omega_{{\boldsymbol{k}}^{\prime}}-{\boldsymbol{k}}\cdot{\boldsymbol{k}}^{\prime}+m^2\right)e^{-\frac{a^2}{2}({\boldsymbol{k}}-{\boldsymbol{k}}^{\prime})^2}\nonumber\\
&-(\omega_{{\boldsymbol{k}}}-\omega_{{\boldsymbol{k}}^{\prime}})^4 \left(\omega_{{\boldsymbol{k}}}\omega_{{\boldsymbol{k}}^{\prime}}-{\boldsymbol{k}}\cdot{\boldsymbol{k}}^{\prime}-m^2\right)e^{-\frac{a^2}{2}({\boldsymbol{k}}+{\boldsymbol{k}}^{\prime})^2}\bigg].
\label{equ16ver1}
\end{align}
Similarly to the canonical formalism, we discard here a divergent temperature-independent vacuum piece. One can easily notice that thermal average values, $\langle:\hat{T}^{00}_{\text{Can},a}:\rangle$ and $\langle:\hat{T}^{00}_{\text{GLW},a}:\rangle$ are same. However, the fluctuations  $\sigma^2_{\text{Can}}(a,m,T)$ and $\sigma^2_{\text{GLW}}(a,m,T)$ are 
different. 

\smallskip
\textit{Hilgevoord-Wouthuysen framework.---}
 Finally, we touch upon another popular choice of the symmetric energy-momentum tensor, i.e., the  Hilgevoord-Wouthuysen form~\cite{HILGEVOORD19631,HILGEVOORD19651002} 
 \begin{align}
\hat{T}^{\mu\nu}_{\text{HW}}&= \hat{T}^{\mu\nu}_{\text{Can}} +\frac{i}{2m} \left(\partial^{\nu}\psbar \sigma^{\mu\beta}\partial_\beta \psi +\partial_\alpha \psbar \sigma^{\alpha\mu}\partial^{\nu}\psi\right) \nonumber\\
&- \frac{i}{4m}g^{\mu\nu} \partial_\lambda \left(\psbar\sigma^{\lambda\alpha}\overleftrightarrow{\partial}_\alpha \psi\right),
\end{align}
where $\sigma_{\mu\nu} \equiv (i/2) \left[\gamma_\mu,\gamma_\nu \right]$. Again, using the Dirac equation and the thermal average of the two and four operators as given by Eqs.~\eqref{equ2ver1} and \eqref{equ4ver1}, we find the thermal average of $\hat{T}^{00}_{\text{HW}}$ and the corresponding fluctuation (discarding again the divergent vacuum contribution)
\begin{align}
\langle:\hat{T}^{00}_{\text{HW},a}:\rangle
& = 4\int\frac{d^3k}{(2\pi)^3}~\omega_{\boldsymbol{k}}f(\omega_{\boldsymbol{k}}) \equiv\varepsilon_{\text{HW}}(T) 
\label{equ18ver1}
\end{align}
and
\begin{align}
&\sigma^2_{\text{HW}}(a,m,T) =  \frac{2}{m^2}\int dK ~dK^{\prime} f(\omega_{{\boldsymbol{k}}})(1-f(\omega_{{\boldsymbol{k}}^{\prime}}))\nonumber\\
&\times \Big[\left(\omega_{{\boldsymbol{k}}}\omega_{{\boldsymbol{k}}^{\prime}}+{\boldsymbol{k}}\cdot{\boldsymbol{k}}^{\prime}+m^2\right)^2\left(\omega_{{\boldsymbol{k}}}\omega_{{\boldsymbol{k}}^{\prime}}-{\boldsymbol{k}}\cdot{\boldsymbol{k}}^{\prime}+m^2\right)\nonumber\\
& ~~~~~~~~~~~~~~~~~~~~~~~~~~~~~~~~~~~~~~~~~\times e^{-\frac{a^2}{2}({\boldsymbol{k}}-{\boldsymbol{k}}^{\prime})^2}\nonumber\\
&~~-(\omega_{{\boldsymbol{k}}}\omega_{{\boldsymbol{k}}^{\prime}}+{\boldsymbol{k}}\cdot{\boldsymbol{k}}^{\prime}-m^2)^2(\omega_{{\boldsymbol{k}}}\omega_{{\boldsymbol{k}}^{\prime}}-{\boldsymbol{k}}\cdot{\boldsymbol{k}}^{\prime}-m^2)\nonumber\\
& ~~~~~~~~~~~~~~~~~~~~~~~~~~~~~~~~~~~~~~~~\times e^{-\frac{a^2}{2}({\boldsymbol{k}}+{\boldsymbol{k}}^{\prime})^2}\Big].
\label{equ19ver1}
\end{align}
From Eqs.~\eqref{equ11ver1}, \eqref{equ15ver1}, and \eqref{equ18ver1} it is clear that the energy densities obtained for different pseudo-gauge choices are the same, i.e., $\varepsilon_{\text{Can}}(T)=\varepsilon_{\text{BR}}(T)=\varepsilon_{\text{GLW}}(T)=\varepsilon_{\text{HW}}(T)$. On the other hand, the fluctuations of $:\hat{T}^{00}_a:$ are in general different for different pseudo-gauge choices. 

Equations~\eqref{equ12ver1}, \eqref{equ16ver1}, and \eqref{equ19ver1} allow us to determine the energy fluctuations of the ``Gaussian" subsystem $S_a$ of the system $S_V$.  The energy density ($\varepsilon$) and its fluctuation ($\sigma$) obtained for a single particle species can be generalized to incorporate degeneracy factors connected with internal degrees of freedom such as  isospin or color charge. If we take into account the degeneracy factors ($g$) then we should replace $\varepsilon\rightarrow g\varepsilon$ and $\sigma^2\rightarrow g\sigma^2$~\cite{Das:2020ddr}. We should mention that spin and particle-antiparticle degrees of freedom are already included in Eqs.~\eqref{equ11ver1}, \eqref{equ12ver1}, \eqref{equ15ver1}, \eqref{equ16ver1}, \eqref{equ18ver1}, and \eqref{equ19ver1}.

\smallskip
\textit{Thermodynamic limit.---}
Before we present our numerical results, it is instructive to consider the thermodynamic limit, i.e., the limiting case of very large system size $a$. 
Since $S_a$ is a subsystem of the system $S_V$, we expect that in the $a\rightarrow \infty$ limit (still with $a^3 \ll V$), quantum fluctuation as obtained here should reduce to the known expression of statistical fluctuation from classical statistical mechanics~\cite{Huang:1987asp}. This can be verified using the Gaussian representation of the three-dimensional Dirac delta function, i.e.
\begin{align}
    \delta^{(3)}({\boldsymbol{k}}-{\boldsymbol{p}})=\lim_{a \to\infty} \frac{a^3}{(2\pi)^{3/2}}e^{-\frac{a^2}{2}({\boldsymbol{k}}-{\boldsymbol{p}})^2}.
\end{align}
This leads us to a pseudo-gauge-independent formula valid in the large $a$ limit
\begin{align}
\sigma_{\text{Can}}^2 & = ~ 
\frac{4~g}{(2\pi)^{3/2} a^3}
\int \frac{d^3{{k}}}{(2\pi)^3}~\omega_{{\boldsymbol{k}}}^2~f(\omega_{{\boldsymbol{k}}}) (1-f(\omega_{{\boldsymbol{k}}}))\nonumber\\
& = \sigma_{\text{GLW}}^2 = \sigma_{\text{HW}}^2  
\label{equ21ver1}
\end{align}
The large $a$ limit of the fluctuation as given in Eq.~\eqref{equ21ver1} can be expressed in terms of the specific heat at a constant volume
\begin{equation} 
c_V = \frac{d\varepsilon}{dT} = \frac{4~g}{T^2} \int \frac{d^3{{k}}}{(2\pi)^3}~\omega_{{\boldsymbol{k}}}^2~f(\omega_{{\boldsymbol{k}}}) (1-f(\omega_{{\boldsymbol{k}}})).
\end{equation}
Therefore, in the large $a$ limit we find, 
\begin{align}
V_a \sigma_n^2
= \frac{T^2 c_V}{\varepsilon^2}
= V \frac{\langle E^2\rangle-\langle E \rangle^2}{\langle E \rangle^2} \equiv V \sigma^2_E,
\label{equ23ver1}
\end{align}
where $V_a = a^3 (2\pi)^{3/2}$ may be identified as the volume of the ``Gaussian'' subsystem $S_a$. In Eq.~\eqref{equ23ver1}, $E$ represents the energy of system $S_V$ and $V\sigma_E^2$ is the normalized energy fluctuation in the system $S_V$~\cite{Huang:1987asp}. Interestingly, Eq.~\eqref{equ23ver1} is also consistent with the result obtained with purely classical arguments~\cite{Mrowczynski:1997mj}.

\begin{figure}[t]
	\includegraphics[scale=0.45]{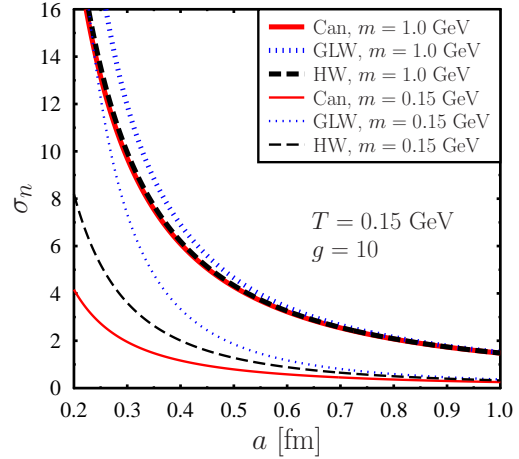}
	\caption{Comparison of normalized standard deviation for various pseudo-gauges for $T=0.15$~GeV, $m=1.0$~GeV (thick lines) and $m=0.15$~GeV (thin lines).}
	\label{fig:1}
\end{figure}
\begin{figure}[t]
	\includegraphics[scale=0.45]{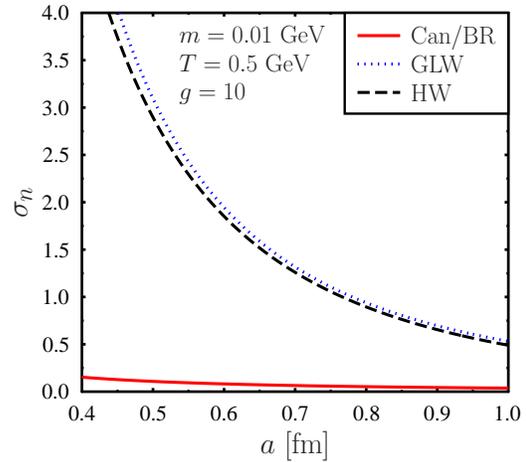}
	\caption{Comparison of normalized standard deviation for various pseudo-gauges for  $T=0.5$~GeV and $m=0.01$~GeV.}
	\label{fig:2}
\end{figure}
%

\smallskip
\textit{Numerical results.---} Our main results describing fluctuations of the energy density in a hot gas of particles with spin $\nicefrac{1}{2}$ are represented by Eqs.~\eqref{equ12ver1}, \eqref{equ16ver1}, and \eqref{equ19ver1}. By straightforward numerical integration, we can obtain from them the results for any subsystem of size $a$, temperature $T$, and particle mass $m$. 

Figure \ref{fig:1} shows a comparison of the normalized standard deviation of fluctuations obtained for three different pseudo-gauges (Can=BR, GLW, HW) for $T=0.15$~GeV. The thick lines correspond to $m=1.0$~GeV, while the thin lines correspond to $m=0.15$~GeV. The internal degeneracy factor is taken to be $g=10$. For $a < 0.5$~fm, we observe that the results obtained with various pseudo-gauges differ, with differences growing as $a$ decreases (although for $m = 1$~GeV, the differences between Can and HW are small). In Fig. \ref{fig:2}, we present our results for small mass, $m=0.01$~GeV, and high temperature, $T=0.5$~GeV. We find that, with growing system size the normalized standard deviation of fluctuations decreases. This behavior may be understood as a consequence of the uncertainty relation. Both Figs.~\ref{fig:1} and \ref{fig:2} also suggest that fluctuations significantly depend on the $m/T$ ratio. With decreasing $m/T$ ratio, fluctuations ($\sigma_n$) decrease, especially in the canonical case.

\begin{figure}[t]
	\includegraphics[scale=0.45]{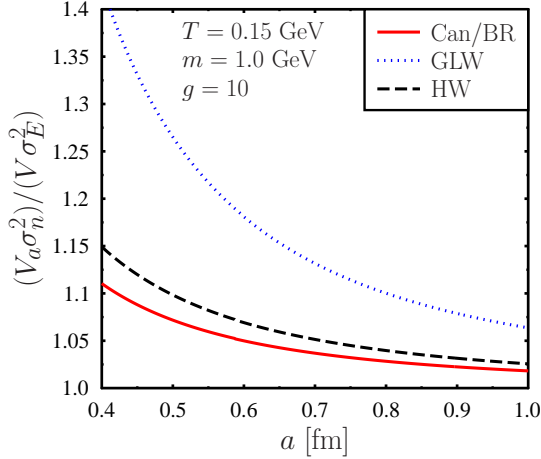}
	\caption{Variation of the normalized energy fluctuation in the subsystem $S_a$ with the length scale $a$ for $T=0.15$ GeV and $m=1.0$ GeV.}
	\label{fig:3}
\end{figure}
\begin{figure}[t]
	\includegraphics[scale=0.45]{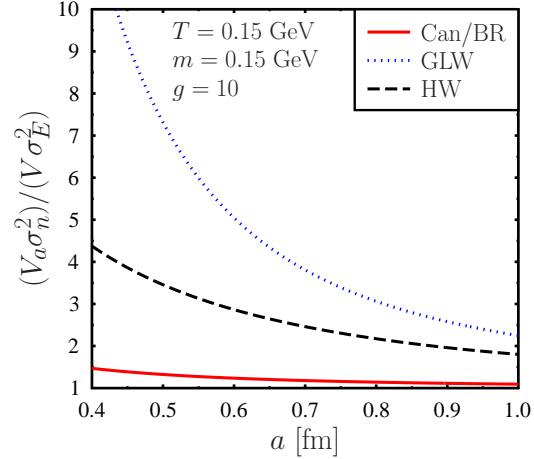}
	\caption{Same as Fig. 3 but for $T=m=0.15$ GeV.}
	\label{fig:4}
\end{figure}

Figures \ref{fig:3} and \ref{fig:4} demonstrate (for $T=0.15$~GeV  and two values of mass, $m=1.0$~GeV and $m=0.15$~GeV, respectively) that the quantum fluctuations of the energy density approach the thermodynamic limit as the system's size increases.  We also see that, with increasing $m/T$ ratio, approach to thermodynamic limit is faster. Interestingly, the smallest fluctuations are found for the canonical pseudo-gauge and in this case the approach to the classical limit is the fastest.

\smallskip
\textit{Conclusions.---} In this work we have delivered explicit expressions for quantum fluctuations of energy density in subsystems of a hot relativistic gas of particles with spin~$\nicefrac{1}{2}$. An intriguing feature of our results is that they depend on the form of the energy-momentum tensor used in the calculations, although for sufficiently large subsystems the results obtained in different pseudo-gauges converge and agree  with the canonical-ensemble formula known from statistical physics. 

As different forms of the energy-momentum tensor of a gas are fundamentally equivalent~\cite{Chen:2018cts}, our finding suggests that the concept of quantum fluctuations of energy is in fact pseudo-gauge dependent. If such fluctuations are measured, they cannot be simply interpreted as a measurement of the fluctuation of a pseudo-gauge independent quantity. What is a priori measured is a particular combination of the field operators, which is different for each pseudo-gauge. The possibility of realistic measurements of such combinations remains an open question that goes beyond the present study. We note that a pseudo-gauge dependence of other physical quantities has been recently also found and discussed in Refs.~\cite{Becattini:2012pp,Nakayama:2012vs}. 

On the practical side, the results of our calculations can be used to determine a scale of coarse graining for which the choice of the pseudo-gauge becomes irrelevant, which may be useful, in particular, in the context of hydrodynamic modeling of high-energy collisions. For any values of $T$ and $m$ we can determine the size $a_0$ such that for $a > a_0$ the fluctuations found in different pseudo-gauges are very close to each other. However, in most of the cases studied in this work this agreement is reached only in the classical limit.

\begin{acknowledgments}
\textit{Acknowledgments.---} We would like to thank Krzysztof Golec-Biernat for very useful and illuminating discussions. This research was supported in part by the Polish National Science Centre Grants No. 2016/23/B/ST2/00717 and No. 2018/30/E/ST2/00432.
\end{acknowledgments}

\bibliography{fluctuationRef.bib}{}
\bibliographystyle{utphys}
\end{document}